\documentclass[aps,preprint,pra]{revtex4}
\usepackage{graphicx}
\usepackage{epsfig,color}
\usepackage{amsmath,amssymb,txfonts}
\usepackage{ulem}

\begin{document}

\title{Remote State Preparation for Quantum Fields}

\author{Ran Ber}
\affiliation{School of Physics and Astronomy, Raymond and Beverly Sackler Faculty of Exact Sciences, Tel Aviv University, Tel-Aviv 69978, Israel.}

\author{Erez Zohar}
\affiliation{Max-Planck-Institut f\"ur Quantenoptik, Hans-Kopfermann-Stra\ss e 1, 85748 Garching, Germany.}

\date{\today}

\begin{abstract}
Remote state preparation is generation of a desired state by a remote observer. In spite of causality, it is well known, according to the Reeh-Schlieder theorem, that it is possible for
relativistic quantum field theories, and a ``physical'' process achieving this task, involving superoscillatory functions, has recently been introduced. In this work we deal with non-relativistic fields, and show that remote state preparation is also possible for them, hence obtaining a Reeh-Schlieder-like result for general fields. Interestingly, in the nonrelativistic case, the process may rely on completely different resources than the ones used in the relativistic case.
\end{abstract}

\maketitle

\section{Introduction}
Remote State Preparation (RSP) \cite{Pati2000,Lo2000,Bennett2001} is a generation of a desired state by a remote observer, who merely acts locally on his own distant system. Such a process is considered successful, if the remote observer is able to make sure that for particular choices of a measurement and its outcome the remote system is in the desired state. Although the success probability might, in general, be small for a single run of the process, it is required that for events with a successful measurement result, the remote state will be arbitrarily close to the desired one (i.e., with fidelity arbitrarily close to 1).

A causal theory is a theory with a maximal velocity $c$, due to which information cannot be transported instantaneously across a distance $R$, but rather at some finite time $T \geq R/c$. Causality is an important property of relativistic systems, and in particular of relativistic quantum field theories (which are usually referred to only as `quantum field theories' - QFTs). As the standard model of high energy physics is a quantum field theory, the importance of causality in our universe and the physical theories describing its behavior could not be overestimated; However, when one wishes to relate relativistic theories, and in particular QFTs, with RSP, he could, naively, argue that the concept of RSP is impossible in a causal theory.

Surprisingly or not, according to an important theorem established long ago by Reeh and Schlieder \cite{Reeh1961,Schlieder1965,Haag1996} in the context of algebraic quantum field theory (AQFT), RSP is possible in relativistic quantum fields. According to the theorem, the action of a field operator, or a polynomial of it, on the vacuum is dense in the Hilbert space - i.e., allowing to generate states which are arbitrarily close to any desired state in the Hilbert space, regardless of its location in configuration space.
Recently \cite{Ber2015}, a physical process resulting with RSP of any desired state was presented, suggesting a ``physical'' way to derive the theorem (rather than AQFT): a prescription for the remote preparation of a desired state was given, rather than a general proof of its existence. This prescription is based on superoscillatory functions, suggesting a deep relation between the task of RSP in QFTs and the mathematical phenomenon of superoscillations \cite{Aharonov1988,Berry1994a}. As the fields are relativistic, it is clear that the resource which allows for such a task is entanglement \cite{Amico2008,Horodecki2009}.
For other quantum fields, which, in general, do not necessarily posses an entangled vacuum state, it remains unclear whether such a task could be accomplished.

One might ask then what is the case for nonrelativistic fields. On one hand, these may be noncausal, and hence one shall not be surprised of noncausal effects such as remote state preparation. On the other hand, how can one be sure that this still allows to remotely prepare any state? In this work we show that this task is possible for a very large class of quantum field theories which satisfy some reasonable physical assumptions. It turns out that RSP is possible, even for fields with a non-entangled vacuum and a bounded group velocity. Thus, a Reeh-Schlieder-like result is obtainable for general quantum fields. However, for fields with a non-entangled vacuum state, the mechanism which produces the remote states relies on an infinite front velocity as a resource, rather than entanglement. This has interesting physical consequences. On one hand, the entanglement-based mechanism, allows generating field states even with a presence of a ``wall'' between the operating region and the target region, as nothing actually propagates in the process. This is not possible in case the mechanism is based on infinite front velocity. On the other hand, the latter mechanism is less sensitive to noises caused by previous attempts to perform the process.

This paper is organized as follows: First, we shall formulate the physical problem we address and describe the process of state preparation in section II. In section III we argue when superoscillations are required and discuss their implications on the success probability. Finally, in section IV, we make the connection to the Reeh-Schlieder theorem and argue that the processes described hereby is a Reeh-Schlieder-like result in the sense that it allows generation of arbitrary field states, even though it may utilise completely different physical resources.

\section{Statement of the problem}
Consider a quantum field theory in $d+1$ dimensions, whose action is invariant under translations and rotations (reflections for $d=1$), but not necessarily under boosts. The dispersion relation of such a theory, due to the rotational invariance, satisfies
\begin{equation}
\omega = \omega\left(k\right),
\label{disp}
\end{equation}
with $k=\left|\mathbf{k}\right|$. Due to the translational invariance, the eigenfunctions take the form
$f_{\mathbf{k}}\left(\mathbf{x},t\right) = h\left(\omega\right) e^{i\left(\mathbf{k}\cdot\mathbf{x} - \omega t\right)}$
and thus the field operator may be expanded as
\begin{equation}
\phi \left(\mathbf{x},t\right) = \int \frac{d^d \mathbf{k}}{\left(2 \pi\right)^d} h\left(\omega_k\right) \left( a_{\mathbf{k}} e^{i\left(\mathbf{k}\cdot\mathbf{x} - \omega_k t\right)}
+ a^{\dagger}_{\mathbf{k}} e^{-i\left(\mathbf{k}\cdot\mathbf{x} - \omega_k t\right)}\right),
\label{field}
\end{equation}
using annihilation and creation operators, satisfying the canonical commutation relation
$\left[a_{\mathbf{k}},a^{\dagger}_{\mathbf{q}}\right] = \left(2 \pi\right)^d \delta ^{\left(d\right)}\left(\mathbf{k}-\mathbf{q}\right)$.
We also assume that $h\left(\omega_k\right)$ is a real function - otherwise, its phase may always be absorbed in the creation and annihilation operators.

The field operator (\ref{field}) is the solution of the Heisenberg equation of motion derived using the field hamiltonian $H_0$. For our purposes, we couple the field to a detector - a two level system, located in $\mathbf{x}=\mathbf{x}_0$, described by the Hamiltonian $H_{\text{d}} = \Omega \sigma_z /2$ ($\hbar=1$). The interaction between the field and the detector is given by
\begin{equation}
H_{\text{int}} = \lambda \left(\sigma_+\epsilon\left(t\right)+\sigma_-\epsilon^*\left(t\right) \right)\phi \left(\mathbf{x}_0\right),
\end{equation}
and is only switched on for a finite period of time - $\epsilon\left(t\right) \neq 0$ only for $t \in \left[-t_0,0\right]$. We assume that in
$t \leq -t_0$ the field is in its vacuum state, and the detector - in its ground state, $\left|\downarrow\right\rangle$. Assuming that $\lambda$ is very small and that $\epsilon\left(t\right)$ is $O\left(1\right)$, one may calculate the state in $t=0$, after the interaction, using first order interaction picture. There
\begin{equation}
H_{\text{int}} = \lambda \epsilon\left(t\right)\sigma_+ e^{i \Omega t} \phi \left(\mathbf{x}_0,t\right) + h.c.,
\end{equation}
and thus the detector-field state after the interaction is given by
\begin{equation}
\left|\Phi,d\right\rangle = \left|0,\downarrow\right\rangle - i \lambda  \int_{-t_0}^{0}dt\epsilon\left(t\right)e^{i \Omega t}
\phi \left(\mathbf{x}_0,t\right)\left|0,\uparrow\right\rangle.
\end{equation}
Post-selecting the spin in the excited state $\left|\uparrow\right\rangle$, one obtains the (normalized) field state
\begin{equation}
\left|\Phi\right\rangle = \mathcal{N}_1^{-1/2} \int \frac{d^d \mathbf{x}}{\left(2 \pi\right)^d} \tilde \epsilon \left(\omega_k+\Omega\right) h\left(\omega_k\right) e^{-i\mathbf{k}\cdot\mathbf{x}} \left|\mathbf{k}\right\rangle,
\label{Phi}
\end{equation}
where $\tilde \epsilon \left(\omega_k+\Omega\right) \equiv \int_{-t_0}^{0}dt\epsilon\left(t\right)e^{i \left(\omega_k+\Omega\right) t}$,
$\mathcal{N}_1 \equiv \int \frac{d^d \mathbf{k}}{\left(2 \pi\right)^d} h^{2}\left(\omega_k\right) \left| \tilde{\epsilon} \left(\omega_k+\Omega\right) \right|^2$ and
$\left|\mathbf{k}\right\rangle \equiv a^{\dagger}_{\mathbf{k}}\left|0\right\rangle$.

Every single excitation of the field may be written as
\begin{equation}
\left|\Psi\right\rangle = \mathcal{N}_2^{-1/2} \int d^d \mathbf{x} F\left(\mathbf{x}\right) \phi\left(\mathbf{x}\right) \left|0\right\rangle,
\label{Psi}
\end{equation}
where
$\mathcal{N}_2 \equiv \int d^d \mathbf{x} d^d \mathbf{y} \frac{d^d \mathbf{k}}{\left(2 \pi\right)^d} h^{2}\left(\omega_k\right) F^{*}\left(\mathbf{y}\right)
e^{i\mathbf{k}\cdot\left(\mathbf{y}-\mathbf{x}\right)}  F\left(\mathbf{x}\right)$ \cite{footnote1}.

We wish to choose a window function $\epsilon \left(t\right)$, such that the state $\left|\Psi\right\rangle$ will be created as a result of the interaction. Thus, we demand that
$\left| \left\langle \Psi | \Phi \right\rangle \right| = 1 $. This is satisfied if we choose
\begin{equation}
\tilde \epsilon_{\text{des}}\left(\omega_k + \Omega\right) = \int d^d \mathbf{x} e^{-i\mathbf{k}\cdot\mathbf{x}} F\left(\mathbf{x}\right),
\label{epsdef}
\end{equation}
where `des' stands for `desired'; i.e., the Fourier transform of the desired window function is equal to the one of $F\left(\mathbf{x}\right)$.

The spherical symmetry restricts us to the creation of states which are symmetric around the detector.
Thus, $F\left(\mathbf{x}\right) = F\left(r\right)$, where $r \equiv \left|\mathbf{x} - \mathbf{x}_0\right|$; WLOG, we assume next that $\mathbf{x}_0=0$.
In this case, the integration reduces to
$\tilde \epsilon_{\text{des}}\left(\omega_k + \Omega\right) = \int dr r^{d-1} F\left(r\right) \int d \theta \sin^{d-2}\theta e^{-ikr\cos\theta} \int d^{d-2} \Omega$. After performing the angular integrals $\int d^{d-2} \Omega = \frac{(d-1) \pi ^{(d-1)/2}}{\Gamma \left((d+1)/2\right)}$ (for derivation see \cite{Flanders1989}) and
$\int d \theta \sin^{d-2}\theta e^{-ikr\cos\theta} = \sqrt{\pi}\left(kr/2\right)^{\left(2-d\right)/2} \Gamma\left(\left(d-1\right)/2\right) J_{\left(d-2\right)/2}\left(kr\right)$
(where $J_{\nu}\left(x\right)$ is a Bessel function), one obtains
\begin{equation}
\tilde \epsilon_{\text{des}}\left(\omega_k + \Omega\right) = \int dr k  F \left( r \right) \left(\frac{2 \pi r}{k}\right)^{d/2} J_\frac{d-2}{2}\left(kr\right).
\label{radialint}
\end{equation}
This equation can be satisfied if and only if $\omega_k = \omega\left(k\right)$ is bijective.
As $r$ increases, the oscillations of the Bessel function (in $k$ space) become faster, hence $\tilde \epsilon_{\text{des}}\left(\omega_k + \Omega\right)$ obtains larger and larger Fourier components. At some point, say $r\geq r_0$, the function $\tilde \epsilon_{\text{des}}\left(\omega_k + \Omega\right)$ begins to have significant Fourier components which oscillate (in frequency space) faster than $t_0$ (below we show that $r_0$ is related to the minimal group velocity of the theory).
Since $\epsilon\left(t\right)$ is nonvanishing only within $[-t_0,0]$, the standard frequency-time relations of Fourier transforms suggest that the above relation could not be satisfied for $r\geq r_0$.

It was shown in \cite{Ber2015} that, for relativistic fields, this problem can be circumvented using a superoscillatory $\tilde \epsilon\left(\omega_k + \Omega\right)$. 
In the following section we introduce this solution briefly. Then, we show that while the same solution holds for non-relativistic fields, some consequences of it are different. This allows us to answer the question ``when are superoscillations required?'' in a general manner.

\section{Superoscillations}
Superoscillatory functions are functions that oscillate faster than their fastest Fourier component \cite{Aharonov1988,Berry1994a}.
This is due to a destructive interference, and thus they are always accompanied by exponentially larger amplitudes somewhere outside the so-called superoscillatory region.
In our implementation, we would have to place the exponentially larger amplitudes in a non-physical domain of $\tilde{\epsilon}$.
Another difficulty regarding superoscillations is that these functions can superoscillate in an arbitrarily large (but not infinite) domain. Therefore, there must also be a physical non-superoscillatory domain.
In our implementation, we shall choose a superoscillatory function that will not be exponentially amplified in this non-superoscillatory domain. Then, we would have to find a way to eliminate the contribution of the function in this domain.

We shall now proceed by finding functions which meet these demands. Consider the following function \cite{Berry1994,Reznik1997,Ber2015}:
\begin{equation}
\tilde{\epsilon}^{[h]}\left(\omega'\right)=\frac{D}{2\delta\sqrt{2\pi}}\int_{0}^{2\pi}d\alpha e^{i\omega't_{0}\left(\frac{\cos\alpha-1}{2}\right)}e^{\frac{i}{\delta^{2}}\cos\left(\alpha-iA\right)},\label{epsilon(omega)-1}
\end{equation}
where $D$, $\delta$ and $A$ are some constants, and $\omega'\equiv\omega+\Omega-\omega_0$ for some $\omega_0$.
Since $t=t_{0}\left(\cos\alpha-1\right)/2$, the function $\epsilon^{[h]}\left(t\right)$ has support only
in $\left[-t_{0},0\right]$. Nevertheless, we shall now prove that
$\tilde{\epsilon}^{[h]}\left(\omega'\right)$ oscillates arbitrarily fast in $\omega'\in\left[0,\omega_{c}\right]$ for $\omega_c\ll\left(\delta^2 t_0 \cosh[A]\right)^{-1}$. Performing the integration explicitly we obtain
\begin{equation}
\tilde{\epsilon}^{[h]}\left(\omega'\right) = \frac{D\sqrt{\pi}}{\sqrt{2}\delta}e^{-\frac{1}{2}i\omega't_{0}} J_{0}\left(\frac{1}{\delta^{2}}\sqrt{1\!+\!\delta^{2}\omega't_{0}\cosh\left[A\right]\!+\!\frac{1}{4}\delta^{4}\omega'^{2}t_{0}^{2}}\right).
\end{equation}
For $\omega'>0$. Using the asymptotic form of the Bessel function \cite{Arfken2012} for $\delta\ll1$, and then taking $\delta^{2}\ll(\omega_{c}t_{0}\cosh\left[A\right])^{-1}$ we obtain
\begin{equation}
\tilde{\epsilon}^{[h]}\left(\omega'\right)\cong De^{-\frac{1}{2}i\omega't_{0}}\cos\left(\frac{1}{\delta^{2}}+\frac{1}{2}\omega't_{0}\cosh\left[A\right]-\frac{\pi}{4}\right)
\end{equation}
in this domain.
Redefining $\tilde{\epsilon}^{[h]}\left(\omega'\right)$ to be the summation of two such functions, one having $\delta^{-2}=2\pi m+\pi/4$, and the other $D \rightarrow \pm iD$ and $\delta^{-2}=2\pi m-\pi/4$, where $m\gg1$, we get
\begin{equation}
\tilde{\epsilon}^{[h]}\left(\omega'\right)=De^{\frac{1}{2}i\omega't_{0}\left(\pm\cosh\left[A\right]-1\right)}.
\end{equation}
This function oscillates in $\omega$ space at ``frequency'' $t'=\frac{1}{2}t_{0}\left(\pm\cosh\left[A\right]-1\right)$.
By increasing $A$ we can set these oscillations to be arbitrarily fast.
The superoscillatory domain is finite, therefore the condition described in Eq. (\ref{radialint}) cannot be  exactly satisfied. However, one can get arbitrarily close to satisfying this condition by increasing the superoscillatory domain. This is achieved by decreasing $\delta$.

Superoscillations come at the price of an exponential growth outside the superoscillatory domain.
In our case the growth occurs at $\omega'<0$. For fields whose energy is bounded from below, we choose $\omega_0$ such that $\omega\geq\omega_{0}$,
therefore $\omega'<0$ corresponds to $\omega+\Omega<\omega_0$, which is in the non-physical domain.
Beyond the superoscillatory domain, the function gradually obtains regular (slower) oscillations, and in the limit $\omega'\gg\omega_{c}$, it behaves like $\omega'^{-1/2}\sin\left(\omega't_{0}\right)$.
In order to eliminate this contribution, one can convolute $\epsilon^{[h]}\left(t\right)$ with some function $h\left(t\right)$ which is differentiable $n$ times ($n\gg1$) and has a small temporal support.

Finally, we use a combination of such superoscillatory functions, each with a different $t'$, in order to generate the window function
\begin{equation}
\tilde{\epsilon}\left(\omega'\right)=\int_{-T}^{T}dt'\tilde{\epsilon}^{[h]}\left(\omega';t'\right)\epsilon_\text{des}\left(t'\right).
\end{equation}
In the limits $T\rightarrow\infty$ and $\delta\rightarrow0$ we get
$\tilde{\epsilon}\left(\omega'\right)\rightarrow\tilde{\epsilon}_{\text{des}}\left(\omega'\right)$
in the segment $\omega'\in\left[0,\omega_{c}\right]$. (This
is while the actual window function, $\epsilon\left(t\right)$, and the desired window function, $\epsilon_{\text{des}}\left(t\right)$, are very different: $\epsilon\left(t\right)$ has temporal support only in $\left[-t_{0},0\right]$, while $\epsilon_{\text{des}}\left(t\right)$ might have an arbitrarily large temporal support.) Therefore, we can generate
remote spherical symmetrical one--particle field states around the spin, up to an arbitrarily small infidelity. The generalization to arbitrary field states is achieved using the same method introduced for relativistic field states \cite{Ber2015}.
Here we only note that the generalization to one--particle states which are not spherical symmetrical involves an array of spins (located at different positions in an arbitrarily small region) rather than a single spin, and the generalization
to many--particle states involves a set of such arrays.

It is shown in \cite{Ber2015} that the exponentially
small amplitude of $\tilde{\epsilon}\left(\omega'\right)$ in the superoscillatory domain results in a success probability of the form
\begin{equation}
P\sim e^{-\frac{\omega_{c}T^2}{t_0}}.
\end{equation}
The finiteness of the superoscillatory domain is responsible for an infidelity
\begin{equation}
\eta\sim\int_{\omega_{c}}^{\infty}\frac{1}{\omega_{c}}\left|\tilde{F}\left(\mathbf{k}\right)\right|^{2}d^{d}\mathbf{k}.
\end{equation}
Since $\tilde{F}\left(\mathbf{k}\right)$ is normalizable, in the limit $\omega_{c}\rightarrow\infty$ one obtains $\eta\rightarrow0$.
Inverting the latter functional relation to $\omega_{c}=\omega_{c}(\eta)\equiv1/g(\eta)$, and expressing $T$ as a functional of the desired state $\left\vert\Psi\right\rangle$, we get the relation
\begin{equation}
P\sim e^{-\frac{T^2\left[\left\vert\Psi\right\rangle\right]}{g\left(\eta\right)t_{0}}}.\label{P as a function of eta}
\end{equation}
When $\tilde{F}\left(\mathbf{k}\right)$ decays with a power law, $g\left(\eta\right)$
behaves according to a  power law as well, and when $\tilde{F}\left(\mathbf{k}\right)$
decays exponentially $g(\eta)\sim1/\text{ln}\left(1/\eta\right)$.
For a given dispersion relation, $T\left[\left\vert\Psi\right\rangle\right]$ can be calculated by plugging $\tilde{\epsilon}_\text{des}\left(\omega_k+\Omega\right)=\int_{-T}^T dt e^{i\left(\omega_k+\Omega\right)t} \epsilon_\text{des}\left(t\right)$ in the l.h.s of Eq. (\ref{radialint}) and expressing the r.h.s as a Fourier transform of a temporal function \cite{footnote2}.

When are superoscillations required? Mathematically, the answer can be deduced from Eq. (\ref{radialint}). However, in order to understand the meaning of that, let us first consider a very simple, non-superoscillatory, window function, $\epsilon \left(t\right) = \delta\left(t+t_0\right)$ - a short and
impulsive interaction, for a quantum field theory in $1+1$ dimensions. Assuming the detector is in the origin, the generated state of the field, after the detector's post-selection, will be
\begin{equation}
\left|\Phi\right\rangle \propto \int_0^{\infty} dk h\left(\omega_k\right) \left(a^{\dagger}_{-k}e^{-i\omega_k t_0} + a^{\dagger}_k e^{-i\omega_k t_0}\right)
\left|0\right\rangle.
\label{deltafield}
\end{equation}
Both terms represent wave packets propagating out of the detector: one is left-moving and the other is right-moving. Assuming that $\omega_k\in\mathbb{R}\,\forall\, k$, and that the group velocity is less than the phase velocity, i.e., $d\omega/dk<\omega/k$, each wave with $k,\omega$ reaches at $t_0$ distance
\begin{equation}
r_0\left(\omega\right) = \frac{\partial \omega}{\partial k}t_0 = v_g \left(\omega\right) t_0,
\label{L0}
\end{equation}
(the minimal $r_0\left(\omega\right)$ is the $r_0$ mentioned above in the simple case of $1+1$ dimensions and a specific desired state). Thus, only waves with frequencies satisfying $v_g \left(\omega\right) t_0 \geq L$ propagate fast enough to arrive to $\pm L$ without superoscillations.
In relativistic theories, as well as other theories in which the group velocity is bounded from above, one can always find $L$ such that all frequencies would require superoscillations, while in theories in which the group velocity is unbounded from above, for every arbitrary $L$ , some frequencies will not require superoscillations.

The above case deals only with waves which are outgoing from the detector. However, one may also consider the case of ingoing wave packets - for example, if one wishes to generate the field state
\begin{eqnarray}
\left|\Psi\right\rangle &=& \left(\phi\left(L\right)+\phi\left(-L\right)\right)\left|0\right\rangle \nonumber\\
&\propto&
\int_0^{\infty} dk h\left(\omega_k\right) \left(\left(a^{\dagger}_{-k} + a^{\dagger}_{k} \right) \left( e^{-i k L} + e^{i k L} \right) \right)
\left|0\right\rangle,
\label{sym}
\end{eqnarray}
which contains both ingoing and outgoing wavepackets. The ingoing waves require $v_g \left(\omega\right) < 0$ for $k \geq 0$, which is impossible, and thus superoscillations are also required for this case.

For general desired states, $L$ is roughly the separation between the operating region and the farthest place in the target region.
Note that the same description holds for higher dimensions; in these cases (when using a single detector) the resulting state is spherically symmetrical. While every single point on the sphere generates waves which propagate in all directions, the spherically symmetrical state as a whole generates waves which propagate only in the $\pm\hat{\bold{r}}$ directions due to the Huygens principle.
Thus, superoscillations are required for the generation of wavepackets which propagate faster than the (slowest) group velocity, or inwards, into the interaction region.

\section{Relation to the Reeh-Schlieder theorem}
It is now the right time to recall a long-standing, possibly surprising result of AQFT - the Reeh-Schlieder theorem \cite{Reeh1961,Schlieder1965,Haag1996}. According to this theorem, the set of Hilbert space vectors $\left|\psi\right\rangle \in \mathcal{H}$ generated from the vacuum (or any other bounded state with a bounded energy \cite{Haag1996}) of a relativistic QFT by operating with polynomials of the field operators in any open region is dense in $\mathcal{H}$.
In other words, by applying certain local operators to the vacuum state in a certain region $O_1$, one is able to generate, with nonzero success probability, a state of the field localized at some remote region(s) $\left\{O_k\right\}_{k\geq 2}$, arbitrarily close to some desired state. The $\left\{O_k\right\}_{k\geq 2}$ regions may remain, throughout the process, outside the light-cone of $O_1$, and thus this outcome must be the result of pre-existing vacuum correlations (the theorem entails a violation of Bell inequalities as well \cite{Clifton1998,Verch2005}).
Consider, as an example, a relativistic scalar field $\phi$ with mass $m$. There, for $\left|\mathbf{x}-\mathbf{x}'\right|^2 \gg m^{-2}$, one obtains the equal-time correlation function
\begin{equation}
\left\langle 0 \right| \phi\left(\mathbf{x},t\right) \phi\left(\mathbf{x}',t\right) \left|0\right\rangle \sim e^{-m \left|\mathbf{x}-\mathbf{x}'\right|},
\label{scalcorr}
\end{equation}
- the correlations do not vanish even between spacelike separated regions
\cite{Peskin1995}.

In this paper, we have shown that RSP is theoretically possible for general quantum fields, including non relativistic ones. Thus, we have described a Reeh-Schlieder-like process for general fields. Note that while the Reeh-Schlieder theorem does not involve time dependence (as one would expect in a relativistic context), our process does. In relativistic theories, adding time dependence is meaningless, as for any two spacelike separated points in spacetime, $x^{\mu} = \left(t,\mathbf{x}\right)$ and $x'^{\mu} = \left(t',\mathbf{x}'\right)$, there is a reference frame (connected by a Lorentz transformation), in which $t=t'$ and $\mathbf{x} \neq \mathbf{x}'$. Thus, if we define $\Delta x^{\mu} \equiv x^{\mu} - x'^{\mu}$, and $r^2 \equiv -\Delta x_{\mu}\Delta x^{\mu} = -\left(t-t'\right)^2 + \left(\mathbf{x}-\mathbf{x}'\right)^2 > 0$, due to Lorentz invariance, in the limit $r^2 \gg m^{-2}$, one obtains $\left\langle 0 \right| \phi\left(\mathbf{x},t\right) \phi\left(\mathbf{x}',t'\right) \left|0\right\rangle \sim e^{-mr}$. In the other extreme case of fields which do not possess any correlations at all - i.e., fields for which
\begin{equation}
\left\langle 0 \right| \phi\left(\mathbf{x},t\right) \phi\left(\mathbf{x}',t\right) \left|0\right\rangle \propto \delta^{\left(d\right)} \left(\mathbf{x}-\mathbf{x}'\right),
\end{equation}
the time dependence is crucial, because correlations are generated in time.
For example, the Schr\"odinger field satisfies
\begin{equation}
\left\langle 0 \right| \phi\left(\mathbf{x},t\right) \phi\left(\mathbf{x}',t\right) \left|0\right\rangle = \delta^{\left(d\right)} \left(\mathbf{x}-\mathbf{x}'\right),
\label{schr}
\end{equation}
and thus, without the time dependence, the overlap between the generated state and the desired state is zero and RSP is not possible.
Therefore, one can deduce that RSP is possible only when
\begin{equation}
\left\langle 0 \right| \phi\left(\mathbf{x},t\right) \phi\left(\mathbf{x}',t'\right) \left|0\right\rangle \neq 0,
\label{diftimes}
\end{equation}
for every $\bold{x}$, $\bold{x}$' and $t \neq t'$. It means that for every quantum field (in which RSP is possible) either there exist correlations in $t=t'$ and/or some components of the quantum field propagate infinitely fast, i.e., the front velocity is infinite.
This is yet another formulation of a result discovered in \cite{Hegerfeldt1974,Hegerfeldt1980} and widely discussed in \cite{Hegerfeldt1994,Petrosky2000}.
In the first scenario, our mechanism uses vacuum correlations as a `resource' for RSP, while in the second scenario the 'resource' is the infinite front velocity. Remarkably, in both cases the (gedanken) prescription for RSP is the same.

The different `resources' have different physical implications. Consider for example a case where one puts a wall between the operating region and the target region. A good model for such a wall, in the case of a scalar field, for example, is a time dependent potential,
\begin{equation}
H_{\text{wall}}=\underset{\Gamma\rightarrow\infty}{\lim} \Gamma\!\!\int \!\!d^d\mathbf{x} W\left(\mathbf{x}\right)\theta\left(t+t_0\right)\phi^{2}\left(\mathbf{x}\right),
\end{equation}
where $\theta\left(t+t_0\right)$ is the Heaviside step function and $W\left(\mathbf{x}\right)=1$ where the wall is placed and $0$ elsewhere.
This potential adds an ``infinite mass'' to the field in particular space points where $W\left(\mathbf{x}\right)=1$, starting from $t=-t_0$ , and thus it acts as a wall.

In causal theories with the equation of motion $\hat{O}\phi\left(x^{\mu}\right) = 0$, the Green's function defined by
\begin{equation}
\hat{O} G\left(x^{\mu}-y^{\mu}\right) = -i\delta^{(4)}\left(x^{\mu}-y^{\mu}\right),
\end{equation}
may be chosen to be a causal, retarded Green's function, $G\left(x^{\mu}-y^{\mu}\right) = D_R\left(x^{\mu}-y^{\mu}\right)$, which vanishes when $y^{\mu}$ is outside the past light-cone of $x^{\mu}$, and thus the field state during the preparation process can be expressed as a superposition of field states at time $t=-t_0$ - all in the past light-cone of the generating region.
Therefore, if the wall is placed outside the past light-cone of both the operating region and the target region throughout the process, its effect, which is propagating at the speed of light at most, will not travel far enough to destroy the vacuum correlations between the two regions and RSP would be possible.

On the other hand, when the process is due to infinite front velocity, the wave front will encounter the wall and RSP will not be possible. Mathematically, this is manifested in the fact that in this case of a non-causal field, there are no light-cone and ``causal'' Green's function. Therefore, the field state $\phi\left(\mathbf{x},-t_0+\epsilon \right)\left\vert 0 \right\rangle$ (where $\epsilon>0$ is arbitrarily small) will involve field states at $t=-t_0$ from regions where the wall is placed.

\begin{figure}
\includegraphics[scale=0.64]{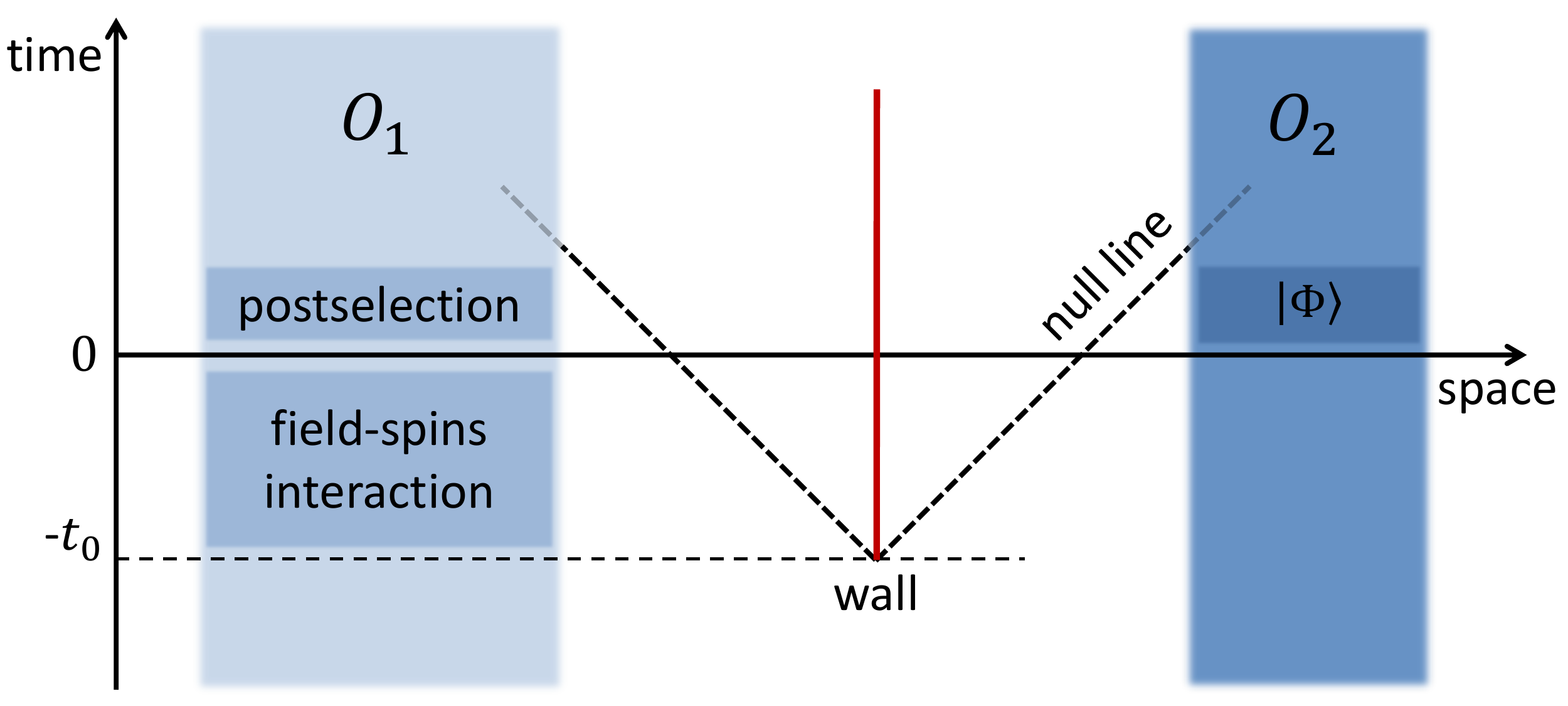}
\caption{A wall which is placed at $-t_0$ between the operating region (denoted as $O_1$) and the target region (denoted as $O_2$) will disrupt the process for fields with an infinite front velocity. However, for relativistic fields, if the wall is placed outside the past light-cone of both $O_1$ and $O_2$ throughout the interaction, the process will not be affected by it.}
\label{remote state generation}
\end{figure}

While the previous example was in favor of the vacuum correlations of relativistic fields, one can also come up with different examples which favor fields with an infinite front velocity. Consider, for example, a case in which one fails to remotely generate a field state and then tries again. The unsuccessful attempt contaminates the vacuum state of the field. Due to high order terms in Eq. (\ref{Phi}), in relativistic fields, the field state will no longer be of bounded energy (i.e. it will be a superposition of energy states with ever increasing energies) \cite{footnote3}. This will destroy the delicate vacuum correlations and so RSP will not be possible. However, for some non-relativistic fields, at least, one could repeat the process on and on. This can be easily demonstrated using fields with a bounded frequency.

In both scenarios, as correlations exist for every $\mathbf{x},\mathbf{x}'$ pair, superoscillations should not be interpreted as the generators of overlap between the initial state and a desired, remote one.
Superoscillations should be rather understood as cancellations of the correlation of the initial state with any other one, except for with the desired state.

Since remote preparation of field states can be used to generate entanglement between distant observers, this result implies that every quantum field
\begin{enumerate}
\item{which is invariant under rotations and translation,}
\item{whose spectrum is bounded from below,}
\item{whose dispersion relation is bijective, and}
\item{whose front velocity is bounded from above}
\end{enumerate}
- possesses a vacuum state which violates Bell's inequalities.

\section{Discussion}

In this paper, we have discussed remote state preparation for non-relativistic fields, and argued that it was possible. This was explicitly shown using a “physical” process.
The scheme that we have considered in this paper is similar to the scheme discussed for relativistic fields in \cite{Ber2015}. It involves coupling the field to a spin (or an array of spins) and then postselecting the spin to a certain state which corresponds to the creation of one (or more) quanta of the field.

In relativistic fields, due to vacuum correlations, particles are never completely localized. I.e., particles created in any arbitrary point in space have a nonvanishing overlap with particles created anywhere else in space. In nonrelativistic fields, although particles may be localized when created, after an infinitesimal time they have a nonvanishing overlap with particles created anywhere else in space as well. I.e., localized particles have an infinite front velocity.
As both vacuum correlations and infinite front velocity induce particles generated in the ``operating region'' to have some (usually very small) overlap with particles generated in the ``target region'', both are mechanisms upon which the process of field RSP may rely.

In order to cancel the overlap between the initial state and states in any other location, except for the target region, a specially tailored window function is used. When the target region is too far for the wave components with the slowest group velocity to reach during the interaction duration, the window function must be superoscillatory. The farther the target region is, the faster the superoscillations have to be. This holds for any theory, either relativistic or not.
The success probability and fidelity of the process depend only on the frequency of the superoscillations, and thus, in particular, they do not depend on the specific mechanism or the existence of a light-cone, but rather on the dispersion relations of the field.
The specific mechanism and the existence of a light-cone become important when the process is slightly altered. For example, when placing a wall between the operating region and the target region before the process begins.

\section*{Acknowledgements}
The authors wish to thank B. Reznik and O. Kenneth for helpful discussions.
EZ acknowledges the support of the Alexander-von-Humboldt foundation.

\bibliography{ref}

\begin{thebibliography}{25}
\bibitem[{foo({\natexlab{a}})}]{footnote1}
\bibinfo{note}{Note that $\mathcal{N}_2$ is, as required, symmetric to exchange
  of $\mathbf{x}$ and $\mathbf{y}$.}
\bibitem[{foo({\natexlab{b}})}]{footnote2}
\bibinfo{note}{For some desired states, a finite $T$ will only be achieved at
  the cost of an additional infidelity. This infidelity could be made
  arbitrarily small.}
\bibitem[{foo({\natexlab{c}})}]{footnote3}
\bibinfo{note}{This feature ``protects'' causality. Without it, one could have
  used this process for signalling.}
\end{thebibliography}

\end{document}